# Radiation hardness study of calorimeter active elements

Victor Kryshkin, Victor Skvortsov

Institute for high energy physics, Protvino, Russian Federation

**Abstract**

There are presented the results of radiation hardness study of calorimeter active elements. One element is based on organic scintillator with wave length shifting fiber light collection. The other one is an element based on a Thick Gas Electron Multiplier (THGEM).

## 1. Introduction

High counting rate experiments require increasing the apparatus radiation hardness. We considered a thick gas electron multiplier as an active element for radiation hard calorimeters [1]. The radiation hardness of a gas detector is defined by radiation hardness of the construction material and gas and aging during irradiation. As the first step we measured the radiation hardness of the construction material of such element. For comparison we also measured the radiation hardness of the active element of the hadron calorimeter that is part of CMS experiment at CERN [2], the element is based on organic scintillator with light collection by a wave length shifting fiber.

## 2. Active elements

A detector element based on a THGEM presented in fig. 1 was produced at IHEP shop using printed circuit board FR-4. The cathode and the anode were manufactured from 0.2 mm thick Cu-clad (35 μm thick) plates. The operating voltages are applied through resistive divider.

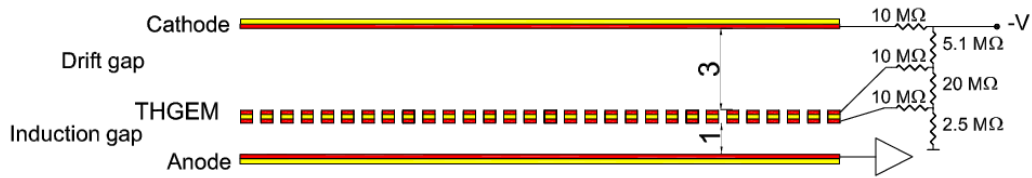

Fig.1 Schematic view of an active element based on THGEM.

Fig. 2 shows THGEM multiplier made of 0.5 mm thick PC board with 35 μm copper clad on both sides. In the octahedral copper electrode there are drilled 0.3 mm holes arranged in a hexagonal pattern with a pitch 0.7 mm; the etched rim around each hole is 0.1 mm.

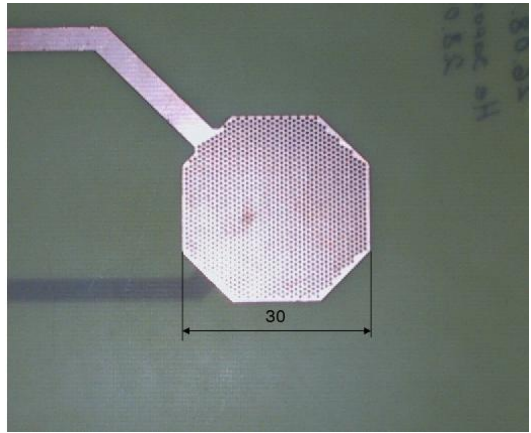

Fig.2 THGEM multiplier.

As one can see on the fig. 3 the distance between the electrodes is fixed by G-10 plates and glued by epoxy. The shape of the cathode and anode is the same as the shape of the THGEM multiplier. Fig. 4 presents the assembled detector.

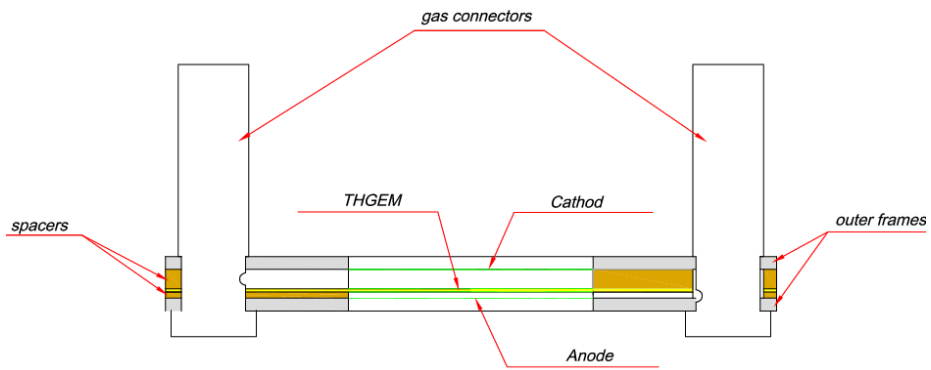

Fig. 3. Schematics of the active THGEM detector.

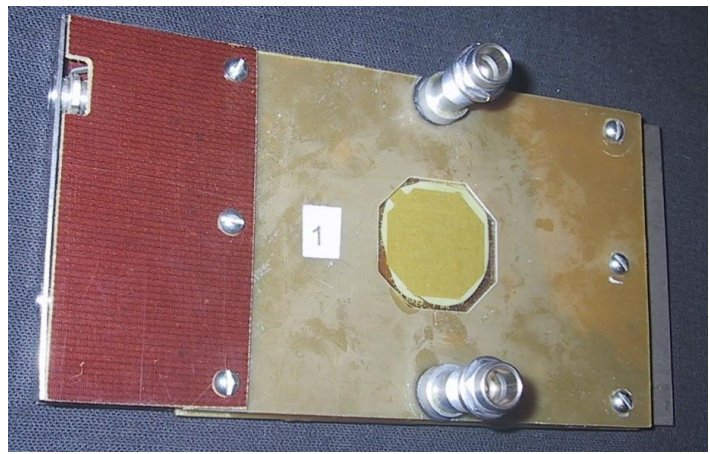

Fig. 4. General view of the THGEM detector.

For comparison we also studied the radiation hardness of the scintillator tile of the end cap CMS hadron calorimeters shown in fig. 5. The element consists of the trapezoidal shape scintillator SCSN-81 [3] 3.7 mm thick with the WLS fiber 1 mm diameter type Y11 produced by Kuraray [4] 20 cm long inserted into the machined groove.

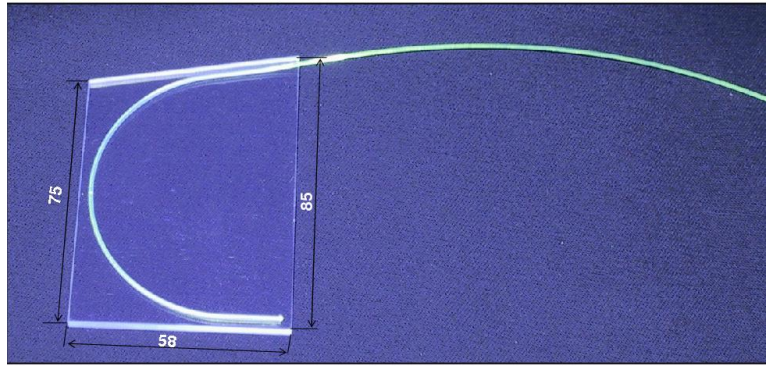

Fig. 5. A photograph of the scintillator tile with WLS fiber.

## 3. Irradiation

Both detectors were irradiated simultaneously – the scintillator was placed behind the THGEM detector - to 5 Mrad by several sources of $^{60}$Co (to provide a uniform exposure to radiation) with intensity 2700 rad/min [5]. The irradiation and the measurements were 3 days apart.

## 4. Measurements

The characteristics of the elements before and after the irradiation were measured in the same conditions. The chamber was flashed by Ar+30%$CO_2$. Simultaneously there was switched on high voltage (2.84 kV), the chamber was exposed to $^{90}$Sr and started the measurements of the current until the current achieved the plateau. Within the measurements error the chamber current was the same before and after the irradiation as well as the pulse height amplitude - about 1 mV on 50 ohm load.

The fiber of the scintillation element was in contact with the photocathode of FEU 85 photomultiplier, in the center of the scintillator there was placed the radioactive source $^{90}$Sr and the current of the phototube was measured at high voltage 1.2 kV. The measurements showed that after the irradiation the current dropped to 34%.

The fig. 6 shows the relative light yield of the CMS hadron calorimeter tile [5] produced of the scintillator SCSN-81 4 mm thick and WLS fiber BCF-91A [6] in dependence on the irradiation dose (points on the dotted line). For comparison we put the values for THGEM (●) and for the scintillator (●).

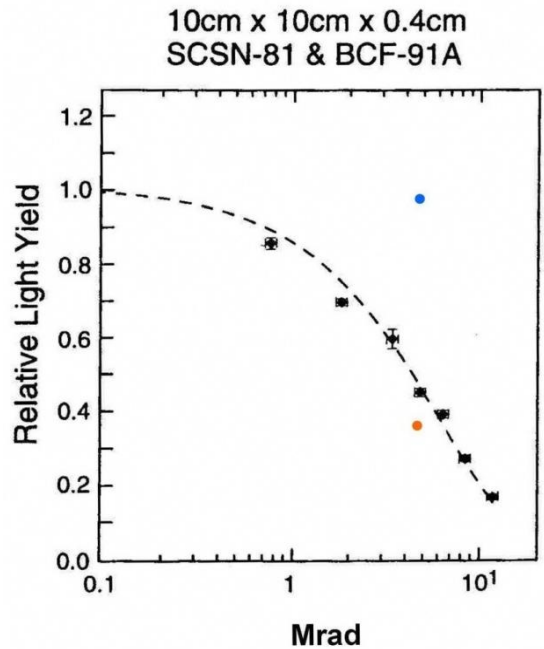

Fig. 6. Relative light yield of the scintillator tile with the light collection by WLS fiber in dependence on dose (points along the dotted line) and the change of response of THGEM (●) and scintillation tile (●) after 5 Mrad irradiation.

## 5. Conclusion

The presented measurements related to the radiation hardness of the THGEM material can be considered as the qualitative evidence that the radiation resistance of the THGEM is very high and the element is a good candidate for detectors operating at high counting rate.

Notice that we used the most available cheap PC boards. More radiation hard materials can be used (in case of need) and the cost will not change appreciably because the cost is mainly defined by the cost of the hole drilling.

As the next step we plan to measure the aging of such detector. To do this we will install a monitor based on THGEM on the proton beam with intensity about $10^9$ per cm$^2$ per second and study its characteristics in time.

**Acknowledgements**